\documentclass[12pt,twoside]{article}
\usepackage{amsmath}
\usepackage{amsfonts}
\usepackage{amssymb}
\usepackage[USenglish,german]{babel}

\newtheorem{thm}{Theorem}[section]

\newtheorem{con}[thm]{Condition}

\newtheorem{prop}[thm]{Proposition}

\newtheorem{lem}[thm]{Lemma}

\newcommand{\rem}{\noindent\textbf{Remark.}\,\,}

\newcommand{\mc}[1]{{\mathcal #1}}

\newcommand{\one}{1}
\def\qed{\relax\ifmmode\hskip2em \Box\else\unskip\nobreak\hskip1em $\Box$\fi}
\newcommand{\Ran}{\mathrm{Ran\,}}
\renewcommand{\Im}{\mathrm{Im\,}}
\renewcommand{\Re}{\mathrm{Re\,}}
\newcommand{\supp}{\mathrm{supp\,}}
\newcommand{\dist}{\mathrm{dist}}
\begin{document}
\selectlanguage{USenglish}

\centerline{\Large{\bf A general resonance theory based on Mourre's 
inequality}}
\begin{center}
Laura Cattaneo${}^{(a)}$, Gian Michele Graf${}^{(b)}$, 
Walter Hunziker${}^{(b)}$
\vspace{0.5cm}

\it  ${}^{(a)}$ Institut f\"ur Angewandte Mathematik,
Universit\"at Bonn,
Wegelerstra\ss e 6,\\
D-53115 Bonn, Germany \\
\it ${}^{(b)}$ Theoretische Physik, ETH Z\"urich, 
CH-8093 Z\"urich,
Switzerland
\end{center}

\begin{abstract}
We study the perturbation of bound states embedded in the continuous spectrum
which are unstable by the Fermi Golden Rule. The approach to resonance theory
based on spectral deformation is extended to a more general class of quantum
systems characterized by Mourre's inequality and smoothness of the
resolvent. Within the framework of perturbation theory it is still possible to
give a definite meaning to the notion of complex resonance energies
and of corresponding metastable states. The main result is a
quasi-exponential decay estimate up to a controlled error of higher order in
perturbation theory. 
\end{abstract}

\section{Introduction and results}
\label{sec:1}

Resonance theory in quantum mechanics deals with the instability of embedded
bound states under arbitrarily weak perturbations. Its task is to provide a
proper foundation for the traditional vague scenario: If 
$H\varphi=\lambda\varphi$ describes the
original bound state $\varphi$, with a simple eigenvalue $\lambda$ embedded 
in the continuous spectrum of the Hamiltonian $H$, and if $H$ is perturbed 
into 
\[H_{\kappa}=H+\kappa V\,,\qquad(\kappa\to 0)\,,\]
then $\lambda$ splits into a complex conjugate pair $\lambda_\kappa$ of 
resonances which dominate
the perturbed evolution of $\varphi$, e.g. as 
$(\varphi,e^{-iH_{\kappa}t}\varphi)\approx e^{-i\lambda_\kappa t}$ up to 
small errors for $t\to+\infty$ and $\Im \lambda_\kappa<0$ (quasi-exponential 
decay). The errors reflect the fact that a strictly exponential
decay is impossible \cite{He} if $H$ is bounded below 
and they may also stem from a perturbative construction
of the resonances $\lambda_\kappa$ such as the time honoured Fermi 
Golden Rule. 

The first consistent resonance theory \cite{Sk1,Sk2,Hu1} along these lines 
required the existence
of an analytic spectral deformation of $H$ which removes the continuous
spectrum of $H$ near $\lambda$ \cite{AgCo1,BaCo1,Si2,RS4,HiSi,HuSi}. 
Meanwhile, Mourre's inequality \cite{Mo1} and related commutator techniques 
\cite{PSS,JMP} have opened a more general and more flexible approach to the
resonance problem \cite{Or2, JeNe}. In this paper we develop one such approach
systematically. Our key result is a quasi-exponential decay law which also
defines the complex resonances $\lambda_\kappa$ uniquely up to subleading
errors. A different, time dependent approach is taken in 
\cite{SW1,MeSi1,CoSo1}. 

We state the basic definitions and assumptions, with comments
on the mathematical background, which will be used freely in the subsequent
parts of the paper. 

The unperturbed quantum system is described by the self-adjoint
Hamiltonian $H$ on the Hilbert space $\mc{H}$. Mourre's operator inequality 
holds for the open interval $\Delta\subset\mathbb{R}$ in the form  
\begin{equation}
E_{\Delta}(H)i[H,A]E_{\Delta}(H)\geqslant \theta E_{\Delta}(H) +K\,,
\label{3_0}
\end{equation}
where $E_{\Delta}(H)$ is  the spectral projection of $H$ for the interval 
$\Delta$, $\theta$ a positive constant and $K$ a compact operator.
$A$ is a self-adjoint operator which needs to be defined in relation to $H$ 
(if possible) for any concrete application, see \cite{CFKS}, Ch.~4 for 
examples. In our general 
setting it suffices to require that the domain of $H$ is invariant under 
the unitary group generated by $A$:
\begin{equation}
e^{isA}\mathcal{D}(H)\subset\mathcal{D}(H)\,,\qquad(s\in\mathbb{R})\,.
\label{1.2}
\end{equation}
This entails the bound 
\begin{equation}
\label{abg}
\|(H+i)e^{isA}(H+i)^{-1}\|\leqslant Ce^{\omega |s|}
\end{equation}
for some constants $C$, $\omega>0 $, see \cite{AmBoGe} Props.~6.3.1 (b) and 
3.2.2 (b), and in turn the fact that the domain 
$\mathcal{D}(H)\cap\mathcal{D}(A)$ is a core for $H$.

At this point we need to comment on the meaning of the commutator $i[H, A]$ of
two possibly unbounded, self-adjoint  operators \cite{JMP}. While the
expression $i (HA-AH)$ is quite useful for casual computations, it is actually
ill defined due to domain questions. The strict definition of this object uses
(\ref{1.2}) and starts 
from the sesquilinear form $i (Hu, Av) - i (Au, Hv)$, which is well defined
for all vectors $u$, $v$ in the domain $\mathcal{D}(H)\cap\mathcal{D}(A)$. 
If this form has a bound 
\begin{equation}
 |i(Hu, Av) - i(Au, Hv)|\leqslant C\|u\|\|(H+i)v\|\,,
\label{bd}
\end{equation}
it extends to the sesquilinear form of a unique self-adjoint operator 
called $i[H, A]$, which is bounded relative
to $H$. Therefore the second order commutator 
$i[i[H, A],A]\equiv -\mathrm{ad}^{(2)}_A(H)$ is defined as well if the bound 
(\ref{bd}) with $i[H, A]$ instead of $H$ on the l.h.s. is assumed. 
The $k$-th order commutator, denoted by $\mathrm{ad}^{(k)}_A(H)$, 
is then defined recursively in terms of $\mathrm{ad}^{(k-1)}_A(H)$ and $A$, 
starting with $\mathrm{ad}^{(0)}_A(H)\equiv H$. For $k=1$ we use 
$\mathrm{ad}^{(1)}_A(H)=[H, A]$ as equivalent notations. 
Multiple commutators appear in connection with resolvent 
smoothness \cite{JMP}: If the commutators 
$\mathrm{ad}^{(k)}_A(H)$, ($k=0,\ldots n+1$) exist, then the weighted 
resolvent and its derivatives up to order $n-1$,
\begin{equation}
\frac{d^k}{dz^k} (A-i)^{-s}(z-H)^{-1}(A+i)^{-s}\,, 
\qquad(k=0,\ldots n-1;\, s>n-1/2 )\,,
\label{bv}
\end{equation}
have one-sided boundary values as $z=x+iy\in\mathbb{C}\setminus\mathbb{R}$ 
approaches the real axis, $\pm y\downarrow 0$, with $x\in\Delta$, 
provided the inequality (\ref{3_0}) holds with $K=0$.

Let $P$ be the eigenprojection corresponding to $\lambda$ and
$\bar{P}$ the projection $\one-P$. Moreover, we write
$\widehat{T}:=\bar{P}T\bar{P}$ for the
restriction of an operator $T$ to the range of $\bar{P}$. We will see 
that 
\[F(z,\kappa)=(\varphi, V\bar{P}(z-\widehat{H}_{\kappa})^{-1}\bar{P}V\varphi)\]
has boundary values as in (\ref{bv}) if $\lambda$ lies in a small Mourre
interval $\Delta$, even though the virial theorem now requires $K\neq 0$ in
(\ref{3_0}). We furthermore assume that the decay rate $\Gamma$ of $\varphi$,
as computed by the prescription of the Fermi Golden Rule, is positive:
\begin{equation}
\frac{\Gamma}{2} := -\Im F(\lambda +i0, 0)>0\,.
\label{agg_4}
\end{equation}
By this condition the eigenvalue $\lambda$ must be embedded in the 
continuous spectrum.

The main result is that there is a resonant state whose time evolution is
consistent with that prescription over a long time interval. Before stating 
it let us summarize the general hypotheses.

\begin{con}
\begin{itemize}
\item [a)] The operator $H$ is self-adjoint and $\lambda$ is an
eigenvalue of $H$, with eigenprojection $P$. The perturbation
operator $V$ is symmetric and $H$-bounded. The Hamiltonian is 
$H_{\kappa}=H+\kappa V$.
\item [b)] 
There is a self-adjoint
operator $A$ and such that (\ref{1.2}) holds true and, for some integer $\nu$, 
the multiple commutators 
$\mathrm{ad}^{(k)}_A(H)$, $\mathrm{ad}^{(k)}_A(V)$, ($k=0,\ldots\nu$) exist as
$H$-bounded operators in the sense explained above. 
\item [c)] Mourre's inequality (\ref{3_0}) holds for some open interval
$\Delta\ni\lambda$. 
\end{itemize}
\label{3_c1}
\end{con}

By the last condition the degeneracy of the eigenvalue $\lambda$ is finite
(see e.g. \cite{CFKS}, Thm. 4.7). 
For simplicity we state the quasi-exponential law only for the case that
$\lambda$ is non-degenerate. 

\begin{thm}\label{thm1}
Let Condition \ref{3_c1} be fulfilled for $\nu \geqslant n+5$, with $\lambda$ 
a simple eigenvalue, as well as eq.~(\ref{agg_4}).
Then there is a function $g\in C^{\infty}_0(\Delta)$ with $g=1$ near $\lambda$
such that
\begin{equation}
(\varphi,e^{-iH_{\kappa}t}g(H_{\kappa})\varphi)
=a(\kappa)e^{-i\lambda_\kappa t}+b(\kappa,t)\,, \qquad (t\geqslant 0)\,,
\label{G_23}
\end{equation}
where 
\begin{equation}\label{remest}
\begin{aligned}
|a(\kappa)-1| &\leqslant  c\kappa^{2}\,,\\
|b(\kappa,t)| &\leqslant
\begin{cases}c\kappa^{2}|\log|\kappa||(1+t)^{-n}\,, \\
c\kappa^{2}(1+t)^{-(n-1)}\,.
\end{cases}
\end{aligned}
\end{equation}
Moreover,
\begin{equation}
\lambda_\kappa=\lambda+\kappa(\varphi, V\varphi)+ \kappa^2F(\lambda
+i0, 0)+o(\kappa^2)\,,
\label{cf}
\end{equation}
and, in particular, $\Im \lambda_\kappa <0$.
\end{thm}

We stress that the quasi-exponential decay (\ref{G_23}) uniquely defines
the resonance $\lambda_\kappa$ up to relative errors $O(\kappa^{4})$. This results
from the following observation and from (\ref{cf}).

\begin{prop}\label{prop0} Assume (\ref{G_23}) with remainder estimates 
(\ref{remest}), say of the second type for $b(\kappa,t)$. If 
\begin{equation}
(\varphi,e^{-iH_{\kappa}t}g(H_{\kappa})\varphi)
=\widetilde{a}(\kappa)e^{-i\widetilde{\lambda}_\kappa t}+\widetilde{b}(\kappa,t)
\label{G_24}
\end{equation}
with similar bounds on $\widetilde{a}$, $\widetilde{b}$, then
\begin{equation}
|\lambda_\kappa-\widetilde{\lambda}_\kappa|\leqslant c\kappa^{2}|\Im \lambda_\kappa| \label{G_25}
\end{equation}
for some constant $c$. 
\end{prop}
 
The proof of Theorem~\ref{thm1} depends on the Feshbach formula \cite{Fe,Li1,
Ho1, Ho2} 
\begin{equation}
(\varphi, (z-H_{\kappa})^{-1}\varphi)^{-1}=z-B(z,\kappa)\,,\qquad
B(z,\kappa)=
\lambda+\kappa(\varphi, V\varphi)+\kappa^2F(z,\kappa)\,,
\label{fesh}
\end{equation}
($\Im z\neq 0$). It furthermore rests on the following results,
which do not require that the eigenvalue $\lambda$ be simple.

\begin{thm}
Let Condition \ref{3_c1} be fulfilled for $\nu\geqslant n+2$. Then 
\[\Ran P\subset\mathcal{D} (A^n)\,.\]
In particular, the operators $A^{n}P$ and $PA^{n}$ are bounded. Moreover,
$\Ran A^{n}P\subset \mathcal{D}(H)$.
\label{t3_1}
\end{thm}

As already mentioned, a second consequence of Condition \ref{3_c1} is the
regularity of the boundary values of the resolvent in an interval
$I\in\Delta$ containing $\lambda$.\\

\begin{thm}\label{thm3} 
Let Condition \ref{3_c1} be fulfilled for $\nu\geqslant n+3$ and let $I$ be a compact subset of the Mourre
interval $\Delta$. For $s>n -\frac{1}{2}$, $\kappa$ small enough, and $z\in I^{\pm}_a :=\{x+iy\vert x\in I,
0<\vert y\vert\leqslant a\}$ define
\begin{equation}
\label{resolvent}
R(z,\kappa) = (A-i)^{-s}
(z-\widehat{H}_{\kappa})^{-1}(A+i)^{-s}\,,
\end{equation}
Then there exist constants $c_1$ and $c_2$ such that
\begin{eqnarray*}
\left\|\frac{d^k}{dz^k}R(z,\kappa) \right\| &\leqslant &
c_1\,,\qquad(k=0,\ldots, n-1)\,,\\
\left\| \frac{d^{n-1}}{dz^{n-1}}R(z,\kappa) -\frac{d^{n-1}}{dz^{n-1}}R(z',\kappa)
\right\| &\leqslant & c_2 \vert
z-z'\vert^{\frac{2s-2n+1}{2s-2n+2s n +1}}\,.
\end{eqnarray*}
Moreover, derivatives and boundary value limits (both in operator norm) may be
interchanged.
\end{thm}
\rem In the case $n=1$ and $\frac{1}{2}<s<1$ the stronger result
\begin{equation}
\| R(z,\kappa) -R(z',\kappa')\|\leqslant c_2 (\vert\kappa -\kappa'\vert +\vert z-z'\vert)^{\frac{2s-1}{2s+1}}
\label{agg_40}
\end{equation}
holds \cite{HuSi}.\\

As we will show in the next section, Theorem \ref{thm3} is a consequence
of results on resolvent smoothness obtained in \cite{JMP} and of
Theorem~\ref{t3_1}. The latter result was proven in \cite{Ca1}, and a somewhat
different proof is given below. The proofs of Theorem~\ref{thm1} and 
Prop.~\ref{prop0} will be given in Section \ref{sec:3}.

\section{Proofs of the preliminary results}
\label{sec:2}

The idea of proof of Theorem~\ref{t3_1} can be traced back to
\cite{FroHe1}. If simplified to the extreme of becoming incorrect, it is as
follows: Consider
\begin{equation}
i[H,A^{2n+1}]=(2n+1)A^ni[H,A]A^n+\,\textrm{lower order terms}\,,
\label{anp1}
\end{equation}
where the lower order terms contain higher order commutators. We take
expectation values in an eigenstate, $H\varphi =\lambda\varphi$, so that the l.h.s. vanishes. As for the r.h.s., we note that
\begin{eqnarray*}
(A^nHA^{-n})A^n\varphi &=& \lambda A^n\varphi\,,\\
A^nHA^{-n} &=& H+\,\textrm{lower order terms.}
\end{eqnarray*}
A diverging vector $A^n\varphi$ would be an approximate eigenstate for $H$, since the lower order terms will become negligible if applied to it. Hence the Mourre estimate is applicable to (\ref{anp1}), so that the r.h.s. diverges as well. This is in contradiction to the l.h.s.\\

We begin with some preliminaries. For $\Im z>0$ we have the representation
$(A-z)^{-1}=i\int_0^\infty e^{-is(A-z)}ds$. Together with a similar
representation for $\Im z<0$, Eq. (\ref{abg}) implies
\begin{equation}
\label{abg1}
\|(H+i)(A-z)^{-1}(H+i)^{-1}\|\leqslant\frac{C}{|\Im z|-\omega}
\end{equation}
for $|\Im z|>\omega$. We will assume this for $z\in\mathbb{C}$, as it will make sure that the following
commutators define bounded operators $\mathcal{D}(H)\to\mathcal{H}$, where $\mathcal{D}(H)$ is equipped with the
graph norm of $H$. Operators identities like the following may then be verified first as forms on
$\mathcal{D}(H)$:
\begin{gather}
\lefteqn{[H,(A-z)^{-1}]=-(A-z)^{-1}{\rm ad}_A(H)(A-z)^{-1}}\hspace{15cm}\label{anp2}\\
=-\frac{1}{2}((A-z)^{-2}{\rm ad}_A(H)+{\rm ad}_A(H)(A-z)^{-2})+ \frac{1}{2}(A-z)^{-2}{\rm ad}_A^{(3)}(H)(A-z)^{-2}\,.\label{anp3}
\end{gather}
We shall need a bounded approximation, $A_\varepsilon = f(A)$, to the unbounded operator $A$, such that $f'(\mu)>0$. The choice
\[A_\varepsilon = \varepsilon^{-1}\arctan(\varepsilon A)\,,\qquad
(\varepsilon\neq 0)\,,\]
will be convenient due to its explicit representation in terms of resolvents:
\begin{equation}
A_\varepsilon = \frac{\varepsilon^{-1}}{2}\int^{\infty}_1[(\varepsilon A +it)^{-1} +(\varepsilon A -it)^{-1}]dt
\label{anp4}
\end{equation}
as a strongly convergent integral. This follows from
\[\arctan\mu=\int^{\infty}_1\frac{\mu}{\mu^{2}+t^2}dt = \frac{1}{2}\int^{\infty}_1\left(\frac{1}{\mu+it}+\frac{1}{\mu-it}\right)dt\,.\]
We remark that for small $\varepsilon$, the resolvents at
$z=\pm i\varepsilon^{-1}t$ appearing in (\ref{anp4}) satisfy (\ref{abg1}).
As suggested by (\ref{anp1}), we will be led to consider $[H,A_\varepsilon]$ as well: by (\ref{anp2}) we have
\begin{equation}
[H,(\varepsilon A+it)^{-1}]=-\varepsilon(\varepsilon A+it)^{-1}{\rm ad}_A(H)(\varepsilon A+it)^{-1}\,,
\label{anp5}
\end{equation}
so that the contributions to $[H,A_\varepsilon]$ from the two terms under the
integral (\ref{anp4}) are now separately convergent in the graph norm of $H$:
\begin{equation}
[H,A_\varepsilon] = -\frac{1}{2}\sum_{\sigma = \pm}\int^{\infty}_1(\varepsilon A+\sigma it)^{-1}{\rm ad}_A(H)(\varepsilon A+\sigma it)^{-1}dt\,.
\label{anp6}
\end{equation}
\begin{lem}
Let $k+l\leqslant\nu$. Then
\begin{equation}
s-\lim_{\varepsilon\to 0}{\rm ad}_{A_\varepsilon}^{(k)}({\rm ad}_A^{(l)}(H))={\rm ad}_A^{(l+k)}(H)\,,
\label{anp7}
\end{equation}
as bounded operators $\mathcal{D}(H)\to\mathcal{H}$.
\label{anp_l1}
\end{lem}
\noindent{\bf Proof.} Clearly, $(\varepsilon A+\sigma
it)^{-1}\mathop{\stackrel{s}{\longrightarrow}}\limits_{\varepsilon\to
0}\,(\sigma it)^{-1}$ in the Hilbert space norm, uniformly in $|t|\geqslant
1$. We claim convergence also as operators $\mathcal{D}(H)\to\mathcal{D}(H)$,
i.e.,
\begin{equation}
s-\lim_{\varepsilon\to 0}(H+i)(\varepsilon A+\sigma it)^{-1}(H+i)^{-1}=
(\sigma it)^{-1}\,.
\label{anp7a}
\end{equation}
In fact, by (\ref{anp5}) we are considering the limit of
\[
(\varepsilon A+\sigma it)^{-1}-\varepsilon(\varepsilon A+it)^{-1}{\rm
ad}_A(H)(\varepsilon A+it)^{-1}(H+i)^{-1}\,
\]
where the second term is bounded in norm by a constant times $\varepsilon$.

We now prove (\ref{anp7}) by induction in $k$. There is nothing to prove for $k=0$. Equation (\ref{anp6}) applies as well to ad$_A^{(l)}(H)$ instead of $H$, showing for $k\geqslant 1$
\begin{eqnarray*}
{\rm ad}_{A_\varepsilon}^{(k)}({\rm ad}_A^{(l)}(H)) &=& {\rm ad}_{A_\varepsilon}^{(k-1)}({\rm ad}_{A_\varepsilon}({\rm ad}_A^{(l)}(H)))\\
&=& -\frac{1}{2}\sum_{\sigma = \pm}\int^{\infty}_1(\varepsilon A+\sigma it)^{-1} {\rm ad}_{A_\varepsilon}^{(k-1)}({\rm ad}_A^{(l+1)}(H))(\varepsilon A+\sigma it)^{-1}dt  \\
&&\hspace{-0.8cm}\mathop{\stackrel{s}{\longrightarrow}}_{\varepsilon\to 0}-\frac{1}{2}\sum_{\sigma = \pm}\int^{\infty}_1(\sigma it)^{-2}dt\cdot{\rm ad}_A^{(l+k)}(H) = {\rm ad}^{(l+k)}(H)
\end{eqnarray*}
in the topology of (\ref{anp7}), where we used (\ref{anp7a}) and the
induction assumption.\qed\\

Another representation for the commutator (\ref{anp6}) is
\begin{eqnarray}
[H,A_\varepsilon] &=& \frac{1}{2}\sum_{\sigma = \pm}(\one-i\sigma\varepsilon A)^{-1}{\rm ad}_A(H)(\one+i\sigma\varepsilon A)^{-1} + \frac{\varepsilon^{2}}{2}(\one+\varepsilon^{2} A^2)^{-1}{\rm ad}_A^{(3)}(H)(\one+\varepsilon^2 A^2)^{-1}\nonumber\\
&&+ \frac{\varepsilon^2}{4}\sum_{\sigma = \pm}\int^{\infty}_1(\varepsilon A+\sigma it)^{-2}{\rm ad}_A^{(3)}(H)(\varepsilon A+\sigma it)^{-2}dt\,.
\label{anp8}
\end{eqnarray}
It is obtained from (\ref{anp3}), rewritten as
\begin{eqnarray*}
[H,(\varepsilon A+it)^{-1}] &=& -\frac{\varepsilon}{2}\left((\varepsilon A+it)^{-2}{\rm ad}_A(H)+{\rm ad}_A(H)(\varepsilon A+it)^{-2}\right)\\
&&+\frac{\varepsilon^3}{2}(\varepsilon A+it)^{-2}{\rm ad}_A^{(3)}(H)(\varepsilon A+it)^{-2}\,,
\end{eqnarray*}
where the last term gives rise to the corresponding one in (\ref{anp8}) through (\ref{anp4}). The contribution of the first term is
\[\frac{1}{4}\sum_{\sigma = \pm}(\one-i\sigma\varepsilon A)^{-1}{\rm ad}_A(H) + {\rm ad}_A(H)(\one-i\sigma\varepsilon A)^{-1}\]
by using $\int^{\infty}_1(\varepsilon A+it)^{-2}dt=-(\one-i\varepsilon A)^{-1}$. This may be regrouped as
\begin{eqnarray*}
\lefteqn{\frac{1}{4}\sum_{\sigma =\pm}(\one-i\sigma\varepsilon A)^{-1}\left({\rm ad}_A(H)(\one+i\sigma\varepsilon A) +(\one-i\sigma\varepsilon A){\rm ad}_A(H)\right)(\one+i\sigma\varepsilon A)^{-1}}\\
&=& \frac{1}{2}\sum_{\sigma =\pm}(\one-i\sigma\varepsilon A)^{-1}{\rm ad}_A(H)(\one+i\sigma\varepsilon A)^{-1} + \frac{\varepsilon}{4}\sum_{\sigma =\pm}i\sigma(\one-i\sigma\varepsilon A)^{-1}{\rm ad}^{(2)}_A(H)(\one+i\sigma\varepsilon A)^{-1}
\end{eqnarray*}
with the first term giving rise to the corresponding one in (\ref{anp8}). Finally, the other term yields the middle one there, since it equals
\begin{eqnarray*}
&&\frac{\varepsilon}{4}(\one+\varepsilon^2 A^2)^{-1}\left(\sum_{\sigma =\pm} i\sigma(\one+i\sigma\varepsilon A){\rm ad}_A^{(2)}(H)(\one-i\sigma\varepsilon A)\right)(\one+\varepsilon^2 A^2)^{-1}\\
&&\hspace{1cm} =\frac{\varepsilon^2}{2}(\one+\varepsilon^2 A^2)^{-1}{\rm ad}_A^{(3)}(H)(\one+\varepsilon^2 A^2)^{-1}\,.
\end{eqnarray*}

The proof of Theorem \ref{t3_1} will depend on a few commutation relations
between $H$ and $A_\varepsilon$. The first one is
\begin{equation}
HA_\varepsilon^n=\sum_{k=0}^n\binom{n}{k}A_\varepsilon^{n-k}{\rm ad}_{A_\varepsilon}^{(k)}(H)\,.
\label{anp9a}
\end{equation}
It is obtained by moving $H={\rm ad}_{A_\varepsilon}^{(0)}(H)$ to the right of $A_\varepsilon$ by means
of
\begin{equation}
{\rm ad}_{A_\varepsilon}^{(j)}(H)A_\varepsilon=A_\varepsilon{\rm ad}_{A_\varepsilon}^{(j)}(H)+{\rm ad}_{A_\varepsilon}^{(j+1)}(H)\,,
\label{anp9b}
\end{equation}
and similarly for the so generated ``contractions'' ${\rm ad}_{A_\varepsilon}^{(k)}(H)$,
($k\geqslant 1$). In this process the number of ways to contract $k$ factors of
$A_\varepsilon$ is $\binom{n}{k}$.
We shall also use the adjoint expansion to (\ref{anp9a}),
\begin{equation}
A_\varepsilon^nH=HA_\varepsilon^n+\sum_{k=1}^n\binom{n}{k}(-1)^k{\rm ad}_{A_\varepsilon}^{(k)}(H)A_\varepsilon^{n-k}\,,
\label{anp10}
\end{equation}
where we singled out the contribution with $k=0$. A further identity is a consequence of (\ref{anp9a},
\ref{anp10}):
\begin{eqnarray}
[H,A_\varepsilon^{2n+1}] &=& [H,A_\varepsilon^{n+1}]A_\varepsilon^n + A_\varepsilon^n[H,A_\varepsilon^{n+1}] - A_\varepsilon^n[H,A_\varepsilon]A_\varepsilon^n\nonumber\\[2mm]
&=& (2n+1)A_\varepsilon^n[H,A_\varepsilon]A_\varepsilon^n\nonumber\\
&& + \sum_{k=2}^{n+1}\binom{n+1}{k}(A_\varepsilon^{n+1-k}{\rm ad}_{A_\varepsilon}^{(k)}(H)A_\varepsilon^n + (-1)^{k-1}A_\varepsilon^{n}{\rm ad}_{A_\varepsilon}^{(k)}(H)A_\varepsilon^{n+1-k})\,,
\label{anp11}
\end{eqnarray}
where we separated the contributions with $k=1$ from the others. The two terms
$k=2$ may be joined to
$A_\varepsilon^{n-1}{\rm ad}_{A_\varepsilon}^{(3)}(H)A_\varepsilon^{n-1}$; the
first one with $k\geqslant 3$ is
\begin{equation*}
A_\varepsilon^{n+1-k}{\rm ad}_{A_\varepsilon}^{(k)}(H)A_\varepsilon^n =A_\varepsilon^{n+2-k}{\rm
ad}_{A_\varepsilon}^{(k)}(H)A_\varepsilon^{n-1} +A_\varepsilon^{n+1-k}{\rm ad}_{A_\varepsilon}^{(k+1)}(H)
A_\varepsilon^{n-1}\,,
\end{equation*}
and similarly for the second. After these steps and except for the term $k=1$,
the highest power of $A_\varepsilon$ flanking a commutator is $n-1$, and the
order of the latter does not exceed $n+2$.\\

\noindent{\bf Proof of Theorem \ref{t3_1}.}
To be shown is that (i) $\varphi\in\mathcal{D}(A^n)$ and (ii)
$A^n\varphi\in\mathcal{D}(H)$ for $\varphi$ with
$H\varphi=\lambda\varphi$. It is convenient to include (iii)
\begin{equation}
\lim_{\varepsilon\to 0}(H+i)A_\varepsilon^n\varphi=(H+i)A^n\varphi
\label{anp11c}
\end{equation}
among the induction assumptions.

Let thus (i-iii) hold true for $k<n$ instead of $n$. From (\ref{anp10})
and (\ref{anp7}) for $l=0$ we obtain, as $\varepsilon\to 0$, that
\begin{gather}
(H-\lambda)A_\varepsilon^n\varphi\,,\label{anp12}\\
(H-\lambda)(\one+i\varepsilon A)^{-1}A_\varepsilon^n\varphi =
(\one+i\varepsilon A)^{-1}(H-\lambda)A_\varepsilon^n\varphi
-i\varepsilon(\one+i\varepsilon A)^{-1}[H,A](\one+i\varepsilon
A)^{-1}A_\varepsilon^n\varphi\label{anp12a}
\end{gather}
are convergent.
For the last statement we used (\ref{anp11c}) for $k=n-1$ and
$(H+i)\varepsilon A_\varepsilon (H+i)^{-1}\stackrel{s}{\longrightarrow} 0$,
which follows from (\ref{anp6}) and
$\varepsilon A_\varepsilon\stackrel{s}{\longrightarrow} 0$.

We observe that (ii, iii) follow once (i) will be established. In fact, we will
then have $A_\varepsilon^n\varphi\to A^n\varphi$, ($\varepsilon\to 0$), so
that (\ref{anp12}) implies (ii, iii) because $H$ is a closed operator.

It thus remains to show (i). Since $(x+i)(x-\lambda)^{-1}$ is bounded for
$x\notin\Delta$, we have
\begin{equation}
\|(H+i)\bar{E}_\Delta(H)(\one+i\varepsilon
A)^{-1}A_\varepsilon^n\varphi\|\leqslant
C\|(H-\lambda)(\one+i\varepsilon A)^{-1}A_\varepsilon^n\varphi\|\leqslant C'\,,
\label{anp13}
\end{equation}
because of (\ref{anp12a}). The expectation of the l.h.s. of (\ref{anp11})
in $\varphi$ vanishes by the virial theorem, whence
\begin{equation}
(2n+1)( A_\varepsilon^n\varphi,i[H,A_\varepsilon]A_\varepsilon^n\varphi)\leqslant C
\label{anp14}
\end{equation}
by (\ref{anp7}). We shall prove that if
\begin{equation}
\|(\one+i\varepsilon A)^{-1}A_\varepsilon^n\varphi\|\longrightarrow\infty
\label{anp15}
\end{equation}
for some sequence $\varepsilon=\varepsilon_n\to 0$, then eventually
\begin{equation}
( A_\varepsilon^n\varphi,i[H,A_\varepsilon]A_\varepsilon^n\varphi)\geqslant\frac{\theta}{2}\|(\one+i\varepsilon A)^{-1}A_\varepsilon^n\varphi\|^2-C_1\|(\one+i\varepsilon A)^{-1}A_\varepsilon^n\varphi\|-C_2
\label{anp16}
\end{equation}
along that sequence. Since this and (\ref{anp14}) are in contradiction with (\ref{anp15}), we have shown
\[\|(\one+i\varepsilon A)^{-1}A_\varepsilon^n\varphi\|\leqslant C\]
for all small $\varepsilon\neq 0$. Hence, for any $\psi\in\mathcal{D}(A^n)$, the l.h.s. of
\begin{equation}
(\psi,(\one+i\varepsilon A)^{-1}A_\varepsilon^n\varphi) = ( (\one-i\varepsilon A)^{-1}A_\varepsilon^n\psi,\varphi)
\label{anp17}
\end{equation}
is bounded by $C\|\psi\|$, and so is its limit $( A^n\psi,\varphi)$. Thus $\varphi\in\mathcal{D}(A^{n*})=\mathcal{D}(A^n)$.\\

To show (\ref{anp16}) we first note that
\[\varphi_\varepsilon:=\frac{(\one+i\varepsilon A)^{-1}A_\varepsilon^n\varphi}{\|(\one+i\varepsilon A)^{-1}A_\varepsilon^n\varphi\|}\stackrel{w}{\longrightarrow} 0\,.\]
Indeed, $\|\varphi_\varepsilon\|=1$ is uniformly bounded and
$(\psi,\varphi_\varepsilon)\longrightarrow 0$ for
$\psi\in\mathcal{D}(A^n)$ by (\ref{anp17}, \ref{anp15}). In particular,
$\|K\varphi_\varepsilon\|\longrightarrow 0$. The contribution to (\ref{anp16}) from the last two terms in (\ref{anp8}) may be estimated from below by
\begin{equation}
-C\varepsilon^2(\|(\one+i\varepsilon A)^{-1}A_\varepsilon^n\varphi\|^2+1)
\label{anp18}
\end{equation}
due to the second part of (\ref{anp13}) (this norm is unaffected by
$i\to -i$). That of the first term is dealt with using
\[{\rm ad}_A(H) = E_\Delta(H){\rm ad}_A(H)E_\Delta(H) + E_\Delta(H){\rm ad}_A(H)\bar{E}_\Delta(H) +\bar{E}_\Delta(H){\rm ad}_A(H)\,,\]
the Mourre estimate $c$) in Condition \ref{3_c1}, and (\ref{anp13}):
\begin{eqnarray*}
\lefteqn{((\one+i\sigma\varepsilon A)^{-1}A_\varepsilon^n\varphi,i{\rm ad}_A(H)(\one+i\sigma\varepsilon A)^{-1}A_\varepsilon^n\varphi)}\\
&\geqslant & \theta\|E_\Delta(H)(\one+i\sigma\varepsilon A)^{-1}A_\varepsilon^n\varphi\|^2 -\|(\one+i\sigma\varepsilon A)^{-1}A_\varepsilon^n\varphi\|\,\|K(\one+i\sigma\varepsilon A)^{-1}A_\varepsilon^n\varphi\|\\
&& -C_1\|(\one+i\sigma\varepsilon A)^{-1}A_\varepsilon^n\varphi\|-C_2\,.
\end{eqnarray*}
By using (\ref{anp13}) once more, $E_\Delta(H)$ may be dropped at the expense of increasing $C_2$. Reducing $\theta$ to $\theta/2$ accounts for (\ref{anp18}) and $K$ at small $\varepsilon$.\qed\\

We now turn to the proof of Theorem~\ref{thm3}, which in the case $n=1$ can
be found in \cite{HuSi}. In \cite{PSS} it has been
proven that under Condition \ref{3_c1} with $K=0$ the weighted resolvent
$(A-i)^{-s}(z-H)^{-1}(A+i)^{-s}$ has continuous boundary
values, and this result was later extended in \cite{JMP} to derivatives 
in $z$ of finite order. The following two lemmas ensure that the hypotheses, 
and in particular $K=0$, are satisfied if the operators ${H}_{\kappa},\,A$
are both restricted to the range of $\bar{P}$.

\begin{lem}\label{lm2.2}
Let Condition \ref{3_c1} be satisfied for $\nu=n+2$. There is a
$\kappa_0$ such that the
commutators
$\mathrm{ad}^{(j)}_{\widehat{A}}(\widehat{H}_{\kappa})$ are
$\widehat{H}_{\kappa}$-bounded for $j=1,2,\dots,n$, uniformly in
$\kappa\leqslant\kappa_0$. \label{rs_l1}
\end{lem}
\noindent{\bf Proof.}
We claim that for $j\leqslant n$
\begin{equation}
\label{hatj}
\mathrm{ad}^{(j)}_{\widehat{A}}(\widehat{H}) =
\bar{P}\mathrm{ad}^{(j)}_{A}(H)\bar{P} +\widehat{G}\,,
\end{equation}
where $G$ is a bounded operator. More precisely, $G$ is a sum of terms
$TB\widetilde{T}^*$, where $B$ is bounded and $T$,
$\widetilde{T}$ are of the form $A^{j_1}\mathrm{ad}^{(j_2)}_{A}(H)P$, with
$j_1+j_2+\tilde j_1+\tilde j_2\leqslant j$. That $G$ is bounded follows
by Theorem~\ref{t3_1} and its pattern is seen to be compatible with induction
from
\begin{gather*}
[\bar{P}\mathrm{ad}^{(j)}_{A}(H)\bar{P},\widehat{A} ]=
\bar{P}\mathrm{ad}^{(j+1)}_{A}(H)\bar{P}+\bar{P}(AP\mathrm{ad}^{(j)}_{A}(H)-\mathrm{ad}^{(j)}_{A}(H)PA)\bar{P}\,,\\
\widehat{A}T=\bar{P}(AT-APT)\,.
\end{gather*}
Similarly, (\ref{hatj}) holds with $V$ in place of $H$. Because of part
(b) of our assumption, i.e.,
\begin{eqnarray*}
\Vert\mathrm{ad}_A^{(j)}(H)\psi\Vert &\leqslant & a^{(1)}_j\Vert H\psi\Vert +
b^{(1)}_j\Vert\psi\Vert\,,\\
\Vert\mathrm{ad}_A^{(j)}(V)\psi\Vert &\leqslant & a^{(2)}_j\Vert H\psi\Vert + b^{(2)}_j\Vert\psi\Vert\,,
\end{eqnarray*}
we have
\begin{eqnarray*}
\Vert\bar{P}\mathrm{ad}^{(j)}_{A}(H_{\kappa})\bar{P}\psi\Vert &\leqslant & \Vert\bar{P}\mathrm{ad}^{(j)}_{A}(H)\bar{P}\psi\Vert + \kappa\Vert\bar{P}\mathrm{ad}^{(j)}_{A}(V)\bar{P}\psi\Vert\\
&\leqslant & (a^{(1)}_j+\kappa a^{(2)}_j)\Vert\widehat{H}\psi\Vert + (b^{(1)}_j+\kappa b^{(2)}_j)\Vert\psi\Vert\,,
\end{eqnarray*}
where
\[\Vert\widehat{H}\psi\Vert\leqslant\Vert\widehat{H}_{\kappa}\psi\Vert +\kappa\Vert\widehat{V}\psi\Vert\leqslant \Vert\widehat{H}_{\kappa}\psi\Vert + a^{(2)}_0\kappa\Vert\widehat{H}\psi\Vert +b^{(2)}_0\kappa\Vert\psi\Vert\,.\]
Thus for $a^{(2)}_0\kappa < 1/2$ the reduced operator $\widehat{H}$ is
$\widehat{H}_{\kappa}$-bounded, and so are the operators $\bar {P}\mathrm{ad}_A^{(j)}(H_\kappa)\bar{P}$.\qed
\begin{lem}
There is an open interval $\Delta$, $\lambda\in\Delta$, and a constant $\kappa_0>0$ such that for $\kappa<\kappa_0$ and some constant $\theta>0$ independent of $\kappa$
\begin{equation}
E_{\Delta}(\widehat{H}_{\kappa})i[\widehat{H}_{\kappa},\widehat{A}]E_{\Delta}(\widehat{H}_{\kappa})\geqslant\theta E_{\Delta}(\widehat{H}_{\kappa})\,,
\label{rs_17}
\end{equation}
where $E(\widehat{H}_{\kappa})$ is the spectral projection of $\widehat{H}_{\kappa}$ on the reduced space $\bar{P}\mc{H}$.
\label{rs_l2}
\end{lem}
\noindent{\bf Proof.} We multiply (\ref{3_0}) from both sides with $\bar{P}$,
commute it with $E_{\Delta}(\widehat{H})$ and $H$, and use that
$E_{\Delta}(\widehat{H})$ converges strongly to $0$ for
$\Delta\to\{\lambda\}$. We so obtain
\[E_{\Delta}(\widehat{H})i[\widehat{H},\widehat{A}]E_{\Delta}(\widehat{H})\geqslant\theta E_{\Delta}(\widehat{H})\]
for some $\theta>0$ and small $\Delta$.\\
\indent Let $h\in C^{\infty}_0(\mathbf{R})$ with $\supp h\subset\Delta$ and $h\vert_{\Delta'}=1$ for some smaller interval $\Delta'$. Then
$E_{\Delta}(\widehat{H})h(\widehat{H})=h(\widehat{H})$
and
\[h(\widehat{H})i[\widehat{H},\widehat{A}]h(\widehat{H})\geqslant\theta h(\widehat{H})\,.\]
Since $h(\widehat{H}_{\kappa})-h(\widehat{H}) = O(\kappa)$ in norm,
\begin{equation}
h(\widehat{H}_{\kappa})i[\widehat{H},\widehat{A}]h(\widehat{H}_{\kappa})\geqslant\theta h(\widehat{H}_{\kappa})-O(\kappa)\,.
\label{rs_1}
\end{equation}
Multiplying (\ref{rs_1}) with $E_{\Delta'}(\widehat{H}_{\kappa})$ we obtain
\begin{equation*}
E_{\Delta'}(\widehat{H}_{\kappa})i[\widehat{H},\widehat{A}]E_{\Delta'}(\widehat{H}_{\kappa})\geqslant(\theta -O(\kappa)) E_{\Delta'}(\widehat{H}_{\kappa})\,.
\end{equation*}
By (\ref{hatj}) we have
\[ [\widehat{V},\widehat{A}]= \bar{P}[V,A]\bar{P} + \widehat{G}\]
with $\widehat{G}$ bounded, and Lemma \ref{rs_l1} implies that $E_{\Delta'}(\widehat{H}_{\kappa})[\widehat{V},\widehat{A}]E_{\Delta'}(\widehat{H}_{\kappa})$ is bounded for $\kappa$ sufficiently small. Therefore
\begin{eqnarray*}
E_{\Delta'}(\widehat{H}_{\kappa})i[\widehat{H}_{\kappa},\widehat{A}]E_{\Delta'}(\widehat{H}_{\kappa}) &\geqslant & E_{\Delta'}(\widehat{H}_{\kappa})i[\widehat{H},\widehat{A}]E_{\Delta'}(\widehat{H}_{\kappa}) - O(\kappa)\\
&\geqslant & (\theta - O(\kappa)) E_{\Delta'}(\widehat{H}_{\kappa})
\end{eqnarray*}
and the statement holds for $\kappa$ small enough. \qed\\

\noindent{\bf Proof of Theorem \ref{thm3}.} By Theorem 2.2 of \cite{JMP} and
the above lemmas we obtain the claim, except that $A$ in (\ref{resolvent}) is
replaced by $\widehat{A}$. Indeed, by \cite{JMP} the existence of
$\mathrm{ad}^{(n+1)}_{\widehat{A}}(\widehat{H}_{\kappa})$ is required, whence
the condition $\nu\geqslant n+3$ through Lemma~\ref{lm2.2}. By induction, we see
that $(\widehat{A})^n=\bar{P}(A^n+G)$, where the bounded operator $G$ is
a sum of terms of the form $A^{n_1}PB$ with $B$ bounded and
$n_1\leqslant n$. We conclude that $(\widehat{A})^n(A+i)^{-n}$ is bounded,
whence Theorem~\ref{thm3} holds as stated. \qed\\

\section{Construction and decay of the metastable state}
\label{sec:3}

The purpose of this section is to prove the nearly exponential decay of a
metastable state as presented in Theorem~\ref{thm1}. Unlike for other results
of this kind we do not assume that the matrix element
$(\varphi, (z-H_{\kappa})^{-1}\varphi)$ has an analytic
continuation across the real axis; instead that quantity, or rather its
inverse, see (\ref{fesh}), will have regular boundary values at real $z$,
as the following lemma shows.

\begin{lem}
Let Condition \ref{3_c1} be fulfilled for $\nu=n+5$, $n\geqslant
0$. On a compact subset $I$ of the Mourre interval $\Delta$,
\[F(E+i0,\kappa)\in C^{n+1}(I)\]
as a function of $E$ whose norm is uniformly
bounded for small $\kappa$:
\begin{equation}
\bigg|\frac{d^j}{dE^j}F(E+i0,\kappa)\bigg|\leqslant c\,,
\qquad(j=0,\ldots, n+1)
\label{agg_l10}
\end{equation}
for some constant $c$. Moreover, 
\begin{equation}
-\Im F(E+i0,\kappa)>0\,,\qquad(E\in I)\,.
\label{agg_l11}
\end{equation}
\end{lem}
\noindent{\bf Proof.} By Theorem \ref{thm3} the claim follows if $(A+i)^sVP$
is bounded for some $s>n+2-\frac{1}{2}$. This is indeed the case for $s=n+2$
thanks to (\ref{anp10}) with $V$ in place of $H$. Eq.~(\ref{agg_l11}) follows
from (\ref{agg_4}) by continuity in
$z$ and $\kappa$, see (\ref{agg_40}), possibly at the expense of making the
interval $I$ smaller.\qed\\

To prove Theorem~\ref{thm1} we will approximate $F(z,\kappa)$ by a function
$F_n(z,\kappa)$ which is analytic across the real axis. Through the Feshbach
formula (\ref{fesh}) there corresponds an approximation $z-B_n(z,\kappa)$ with
\begin{equation}
B_n(z,\kappa)=\lambda_1(\kappa)+\kappa^{2}F_n(z,\kappa)\,,\qquad
\lambda_1(\kappa)=\lambda+\kappa(\varphi, V\varphi)
\label{bn}
\end{equation}
to the (inverse) matrix element
$(\varphi, (z-H_{\kappa})^{-1}\varphi)^{-1}$. This will allow
to compute the survival amplitude (\ref{G_23}) as in the deformation analytic
case, up to a small error consistent with the remainder estimate
(\ref{remest}). We first discuss the requirements for $F_n$:

\begin{prop}
Let $F_n(z,\kappa)$ be an approximation to $F(z,\kappa)$ in the sense that
for $\kappa$ small enough
\begin{enumerate}
\item the function $F_n(z,\kappa)$ is analytic in a neighborhood
$U_r(\lambda)=\{z\in\mathbb{C}|\,|z-\lambda|<r\}$ of $\lambda$ for
some $r>0$ and
\begin{gather}
|F_n(z,\kappa)|\leqslant c\,,\label{fn} \\
-\Im F_n(z,\kappa)> 0\,;\label{imf}
\end{gather}
\item the remainder
\begin{equation}
r_n(E,\kappa):=\left((B(z,\kappa)-z)^{-1}-(B_n(z,\kappa)-z)^{-1}\right)\Big|_{z=E+i0}\,,
\label{agg_72}
\end{equation}
satisfies
\begin{equation}
\|r_n^{(j)}(\cdot,\kappa)\|_{L^{1}(I)}\leqslant
\begin{cases}
c\kappa^{2}\,,\qquad&(j=0,\dots,n-1)\,,\\
c\kappa^{2}|\log|\kappa||\,,\qquad&(j=n)\,.
\end{cases}
\label{rn}
\end{equation}
\end{enumerate}
Then (\ref{G_23}) holds for a suitable $g\in C^{\infty}_0(I)$ and a complex
frequency $\lambda_\kappa$ satisfying
\begin{equation}
\lambda_\kappa=\lambda_1(\kappa)+ \kappa^2F_n(\lambda_1(\kappa),\kappa)+O(\kappa^4)\,.
\label{cf1}
\end{equation}
\label{G_t4}
\end{prop}

We will later show that the $n$-th order Taylor polynomial of $F$ 
at the first order eigenvalue $\lambda_1(\kappa)$ qualifies for 
$F_n(z,\kappa)$. Then
$F_n(\lambda_1(\kappa),\kappa)=F(\lambda_1(\kappa)+i0,\kappa)=F(\lambda
+i0, 0)+o(1)$ as $\kappa\to 0$, and (\ref{cf1}) implies (\ref{cf}),
thus completing the proof of Theorem~\ref{thm1}.\\

To prove the above proposition we need the following lemma.

\begin{lem}
For $\kappa$ small enough the function
$B_n(z,\kappa)-z$ has exactly one zero, $\lambda_\kappa$, in $U_r(\lambda)$. It
satisfies $\Im \lambda_\kappa\leqslant0$ for $\kappa\neq 0$ and (\ref{cf1}). \label{G_l4}
\end{lem}
\noindent{\bf Proof.} Eq.~(\ref{fn}) implies
\begin{equation}
|(B_n(z,\kappa)-z)-(\lambda_1(\kappa)-z)|=\kappa^{2}|F_n(z,\kappa)|\leqslant
c\kappa^{2}\,. \label{G_21}
\end{equation}
It suffices to show this in any neighborhood
$U_{r'}(\lambda)$ with $r'<r$. On
$\partial U_{r'}(\lambda)=\{z\in \mathbb{C}\,|\,|\lambda-z|=r'\}$ we have
$|\lambda_1(\kappa)-z|\geqslant
r'-|\kappa(\varphi,V\varphi)|$, which for small $\kappa$ is bigger
than $c\kappa^{2}$. Therefore,
\[|(B_n(z,\kappa)-z)-(\lambda_1(\kappa)-z)|< |\lambda_1(\kappa)-z|\,,\]
so that by Rouch\'{e}'s theorem, see e.g. \cite{Con}, $B_n(z,\kappa)-z$ and $\lambda_1(\kappa)-z$ have the same
number of zeros in $U_{r'}(\lambda)$, namely one (called $\lambda_\kappa$). It can not lie in the upper half-plane
since $\Im(B_n(z,\kappa)-z)< 0$ there. It can also not lie in $|z-w(\kappa)|\geqslant C\kappa^4$ for sufficiently
large $C$, where $w(\kappa)$ is the expanded part on the r.h.s. of (\ref{cf1}). Indeed,
\begin{multline*}
|B_n(z,\kappa)-z|\geqslant |w(\kappa)-z|-
\kappa^2|F_n(z,\kappa)-F_n(\lambda_1(\kappa),\kappa)|\\
\geqslant |w(\kappa)-z|-C\kappa^2|z-\lambda_1(\kappa)| \geqslant (1-C\kappa^2)|z-w(\kappa)|-
C\kappa^2|w(\kappa)-\lambda_1(\kappa)|\,,
\end{multline*}
since (\ref{fn}) implies a uniform bound on $dF_n/dz$. Given that $|w(\kappa)-\lambda_1(\kappa)|\leqslant c\kappa^2$
the claim follows.
\qed\\

We will next establish Proposition \ref{G_t4} for that complex frequency $\lambda_\kappa$.\\

\noindent{\bf Proof of Proposition \ref{G_t4}.} We choose $g$ such
that $\supp g\subset I$ and $g\equiv 1$ on some smaller interval. We then have
\begin{eqnarray*}
(\varphi, e^{-iH_{\kappa}t}g(H_{\kappa})\varphi)
&=&-\lim_{\varepsilon\downarrow 0}\frac{1}{\pi}(\varphi, \int_{\mathbf{R}}d\mu e^{-i\mu t}g(\mu)\Im (\mu+i\varepsilon-H_{\kappa})^{-1}\varphi)\\
&=&\frac{1}{\pi}\int_{\mathbf{R}}d\mu\, e^{-i\mu
t}g(\mu)\Im (B(\mu+i0,\kappa)-\mu)^{-1}
\end{eqnarray*}
since by (\ref{agg_l11}) the limit can be taken under the integral. We split the expectation value as
\begin{equation*}
(\varphi, e^{-iH_{\kappa}t}g(H_{\kappa})\varphi) =
\frac{1}{\pi}\int_{\mathbf{R}}\!\!d\mu\, e^{-i\mu
t}g(\mu)\Im (B_n(\mu,\kappa)-\mu)^{-1}
+\frac{1}{\pi}\!\!\int_{\mathbf{R}}\!\!d\mu e^{-i\mu
t}g(\mu)\Im r_n(\mu,\kappa)\,.\label{G_22}
\end{equation*}
Using
\begin {equation}
e^{-i\mu t}=(1+t)^{-n}\left(1+i\frac{d}{d\mu}\right)^{n}e^{-i\mu t}
\label{partint}
\end{equation}
and $g(\mu)=0$ for $\mu\notin I$, the second integral can be estimated
by $(1+t)^{-n}$ times
\begin{multline*}
\bigg|\int_{\mathbf{R}}d\mu\, g(\mu)\Im r_n(\mu,\kappa)\bigg(1+i\frac{d}{d\mu}\bigg)^{n}e^{-i\mu t}\bigg|\\
=\bigg|\int_{\mathbf{R}}d\mu\, e^{-i\mu t}
\bigg(1-i\frac{d}{d\mu}\bigg)^{n}(g(\mu)\Im r_n(\mu,\kappa))\bigg|\leqslant
\max_{0\leqslant j\leqslant n}\|r_n^{(j)}(\cdot,\kappa)\|_{L^{1}(I)}\,,
\end{multline*}
or with $n-1$ in place of $n$. Because of (\ref{rn}) this is consistent with
the remainder estimate for $b(\kappa,t)$ in (\ref{G_23}). The first integral
equals
\begin{equation}
\frac{1}{2\pi i}\int_Id\mu\, e^{-i\mu t}g(z)
\Bigl((B_n(\mu,\kappa)-\mu)^{-1}-(\overline{B_n(\bar{\mu},\kappa)}-\mu)^{-1}
\Bigr)
\label{firstint}
\end{equation}
with both terms in parentheses being analytic functions for
$\mu\in U_r(\lambda)$. Within a smaller interval, where $g(\mu)=1$, we
deform $I$ to a path $\gamma\subset U_r(\lambda)$ in the lower half-plane
staying a positive distance away from $\lambda$. In doing so we cross
the simple pole $\lambda_\kappa$ of the first term (but not $\bar \lambda_\kappa$
of the second). We so obtain for (\ref{firstint})
\begin{equation}
\frac{1}{2\pi i}\int_{\gamma}dz\, e^{-iz t}g(z)\Bigl((B_n(z,\kappa)-z)^{-1}-(\overline{B_n(\bar{z},\kappa)}-z)^{-1}\Bigr)+ e^{-i\lambda_\kappa
t}\mathrm{Res}_{\substack{z=\lambda_\kappa}}(B_n(z,\kappa)-z)^{-1}\,,
\label{G_26}
\end{equation}
where $g$ was extended to $g(z)=1$ along the deformed portion of
$\gamma$. The residue is
\[\mathrm{Res}_{\substack{z=\lambda_\kappa}}(B_n(z,\kappa)-z)^{-1}=\frac{1}{2\pi i}\int_{|z|=r'}dz\,(B_n(z,\kappa)-z)^{-1}=1+O(\kappa^{2})\,,\]
since by (\ref{G_21}) we have $(B_n(z,\kappa)-z)^{-1}= (\lambda_1(\kappa)-z)^{-1} +O(\kappa^{2})$ on $|z|=r'$. The
residue contribution in (\ref{G_26}) thus matches the term $a(\kappa)e^{-i\lambda_\kappa t}$ in (\ref{G_23}). The line
integral on $\gamma$ may be written as
\begin{equation*}
-\frac{\kappa^{2}}{2\pi i}\int_{\gamma}dz\, e^{-iz
t}g(z)(B_n(z,\kappa)-z)^{-1}(F_n(z,\kappa)-\overline{F_n(\bar{z},\kappa)})(\overline{B_n(\bar{z},\kappa)}-z)^{-1}\,,
\end{equation*}
where the integrand, except for $e^{-iz t}$, has derivatives of any order which
are bounded uniformly in small $\kappa$. Using again (\ref{partint}) with
$\mu=z$, repeated integrations by parts, and $|e^{-iz t}|\leqslant1$ for
$t\geqslant 0,\, z\in\gamma$ that integral can thus be lumped into the remainder
$b(\kappa,t)$.\qed\\

As anticipated we will now show that
\begin{equation*}
F_n(z,\kappa):=
\sum_{k=0}^n\frac{1}{k!}\left(\frac{d^k}{dE^k}F(E+i0,\kappa)\left|_{E=\lambda_1(\kappa)}\right.\right)
(z-\lambda_1(\kappa))^k
\end{equation*}
satisfies the requirements of Proposition \ref{G_t4}. Hypothesis 1 holds by
(\ref{agg_l10}, \ref{agg_l11}) provided $r$ is small enough. As a preliminary
in checking hypothesis 2 we show that
\[G(E,\kappa):= (B(E+i0,\kappa)-E)^{-1}\]
may be bounded by a simpler function,
\[\widehat{G}(E,\kappa):= (\lambda_1(\kappa)-E+i\kappa^2\Im F(\lambda+i0,\kappa))^{-1}\,.\]
\begin{lem}
\begin{equation}
|G(E,\kappa)|\leqslant C|\widehat{G}(E,\kappa)|\,,
\qquad(E\in I)\,. \label{14}
\end{equation}
\end{lem}
\noindent {\bf Proof.} Inequality (\ref{14}) is equivalent to
\begin{equation}
|\lambda_1(\kappa)-E+i\kappa^2\Im F(\lambda+i0,\kappa)|^2\leqslant
C^2|\lambda_1(\kappa)-E+\kappa^2F(E+i0,\kappa)|^2 \label{15}
\end{equation}
and follows since the left hand side equals
\begin{multline*}
  (\lambda_1(\kappa)-E)^2 + (\kappa^2\Im F(\lambda+i0,\kappa))^2\\ \leqslant
2(\lambda_1(\kappa)-E+\kappa^2\Re F(E+i0,\kappa))^2
+2(\kappa^2\Re F(E+i0,\kappa))^2
+(\kappa^2\Im F(\lambda+i0,\kappa))^2\,.
\end{multline*}
By Eqs. (\ref{agg_l10}) and (\ref{agg_l11}) the last two terms are bounded by
a constant times $(\kappa^2\Im F(E+i0,\kappa))^2$ for any $E\in I$. This
proves (\ref{15}). \qed\\

We recall the definition (\ref{bn}) of $B_n(z,\kappa)$ and set
$G_n(z,\kappa)=(B_n(z,\kappa)-z)^{-1}$.
We note that (\ref{agg_l11}) also holds for $F_n$ provided $I$
is small enough (but independent of $\kappa$). Therefore we also
have
\begin{equation}
|G_n(E,\kappa)|\leqslant C|\widehat{G}(E,\kappa)|\,,
\qquad(E\in I)\,. \label{16}
\end{equation}
By Taylor's estimate,
\[|B^{(k)}(E+i0,\kappa)-B_n^{(k)}(E,\kappa)|\leqslant C\kappa^2|E-\lambda_1(\kappa)|^{n+1-k}\]
for $E\in I$, $k=0,\dots,n$, and, since $I$ is bounded, also
\begin{equation}
|B^{(k)}(E+i0,\kappa)-B_n^{(k)}(E,\kappa)|\leqslant
C\kappa^2|E-\lambda_1(\kappa)|^{m} \label{17}
\end{equation}
for any $m\leqslant n+1-k$.
We shall also need that the derivatives of $G_n$ essentially
behave as if $B_n(E,\kappa)$ had no dependence on $E$.
\begin{lem} For $k=0,\dots,n$,
\begin{equation}
\sup_{E\in
I}\left|G_n^{(k)}(E,\kappa)(B_n(E,\kappa)-E)^{k+1}\right|\leqslant
C\,. \label{18}
\end{equation}
\end{lem}
\noindent {\bf Proof.} We shall prove this by induction starting
with $k=0$, which holds by
\[G_n(E,\kappa)(B_n(E,\kappa)-E)=1\,.\]
Taking the $k$-th derivative thereof and multiplying with
$(B_n-E)^k$, we get
\[\sum^k_{j=0}\binom{k}{j}G_n^{(j)}(E,\kappa)(B_n(E,\kappa)-E)^{j+1}\cdot
(B_n(E,\kappa)-E)^{k-j-1} \frac{d^{k-j}}{dE^{k-j}}(B_n(E,\kappa)-E)=0\,.\]
The term $j=k$ is the one under estimate (its second factor equals 1 in this case). Since $j<k$ for the others, their first factors are bounded by induction assumption, and the second manifestly. \qed\\

We can now estimate
$r_n^{(k)}(E)=G^{(k)}(E+i0,\kappa)-G_n^{(k)}(E,\kappa)$.
It is convenient to consider the following, slightly stronger
statement $\mathcal{P}_k$:
\begin{prop}
Let $l\geqslant 0$. Then
\[\left\|(G^{(k)}-G_n^{(k)})\widehat{G}^l\right\|_{L^1(I)}\leqslant
\begin{cases}
C\kappa^2\, \qquad&(l+k<n)\,, \\
C\kappa^2|\log|\kappa||\, \qquad& (l+k=n)\,.
\end{cases}
\]
\end{prop}
We remark that for $l=0$  this estimate is (\ref{rn}).\\

\noindent {\bf Proof.} Assuming
$(\mathcal{P}_0,\dots,\mathcal{P}_{k-1})$ we shall prove
$\mathcal{P}_k$, as long as $k\leqslant n$ (note that the
induction assumption is empty for $k=0$). We differentiate
$G(E,\kappa)(B(E,\kappa)-E)=1$ $k$ times:
\[\sum_{j=0}^k\binom{k}{j}G^{(j)}(E,\kappa)(B(E,\kappa)-E)^{(k-j)}=\delta_{k0}\,.\]
Subtracting from it the same relation for $G_n$, $B_n$ we obtain
\begin{eqnarray*}
\sum_{j=0}^k\binom{k}{j}\hspace{-0.7cm}&&\left[ (G^{(j)}(E,\kappa)-G_n^{(j)}(E,\kappa))(B(E,\kappa)-E)^{(k-j)}\right.\\
&&\quad + \left.
G_n^{(j)}(E,\kappa)(B^{(k-j)}(E,\kappa)-B_n^{(k-j)}(E,\kappa))\right]
=0\,.
\end{eqnarray*}
We then separate the first term for $j=k$ and multiply by
$G\widehat{G}^l$:
\begin{eqnarray*}
\lefteqn{(G^{(k)}(E,\kappa)-G_n^{(k)}(E,\kappa))\widehat{G}^l(E,\kappa)}\\
&&= -\sum_{j=0}^{k-1}\binom{k}{j}(G^{(j)}(E,\kappa)-G_n^{(j)}(E,\kappa))(B(E,\kappa)-E)^{(k-j)}G(E,\kappa)\widehat{G}^l(E,\kappa) \\
&&\quad
-\sum_{j=0}^k\binom{k}{j}G_n^{(j)}(E,\kappa)(B^{(k-j)}(E,\kappa)-B_n^{(k-j)}(E,\kappa))G(E,\kappa)\widehat{G}^l(E,\kappa)\,.
\end{eqnarray*}
Since $B$ and its derivatives are bounded and because of (\ref{14}), the terms of the first sum (which is empty for $k=0$) may be estimated by $|(G^{(j)}-G_n^{(j)})(E,\kappa)\widehat{G}^{l+1}(E,\kappa)|$. For $l+k<n$ its $L^1(I)$-norm is bounded by $C\kappa^2$ by induction assumption, since $(l+1)+j\leqslant (l+1)+(k-1)<n$. If $l+k=n$ it is bounded by $C\kappa^2|\log|\kappa||$ for the same reason.
In view of (\ref{14}, \ref{16}, \ref{17}, \ref{18}) the terms of the
second sum are bounded by a constant times
\begin{equation}
\kappa^2|\widehat{G}|^{j+1}|\widehat{G}|^{l+1}|E-\lambda_1(\kappa)|^m
\label{19}
\end{equation}
for any choice of $m\leqslant n+1-(k-j)$. If $l+k<n$, and hence
$j+l+2\leqslant j+n-k+1$, we may pick $m=j+l+2$, so that
(\ref{19}) is bounded by $\kappa^2$ since
$|\widehat{G}|^{m}|E-\lambda_1(\kappa)|^m\leqslant1$.
If $l+k=n$ we need to take $m$ one less than before, so that
(\ref{19}) is bounded by $\kappa^2|\widehat{G}(E,\kappa)|$. Since
$\|\widehat{G}\|_{L^1(I)}\leqslant C|\log|\kappa||$, the claim again
follows.\qed\\

We conclude with the proof of Prop.~\ref{prop0}, which is independent of
Theorem~\ref{thm1}.
Together with that theorem, it shows
that the resonance energy $\lambda_\kappa$ is uniquely defined by the decay
law (\ref{G_23}) up to errors $O(\kappa^{4})$.\\

\noindent{\bf Proof of Proposition \ref{prop0}.} We will make use of the 
estimate
\begin{equation}
\bigl|\log\frac{w_1}{w_2}\bigr|
\leqslant\frac{2+\pi}{2}\frac{|w_1-w_2|}{\min(|w_1|,|w_2|)}\,,
\qquad(w_1,w_2\in\mathbb{C}\setminus\{0\})\,,
\label{w1w2}
\end{equation}
where the modulus on the l.h.s. should be understood as
$|z|=\dist(z, 2\pi i \mathbb{Z})$, in view of the many values of the
logarithm. We shall prove it later and use it now for $w_1=e^{-i\lambda_\kappa t}$,
$w_2=e^{-i\widetilde{\lambda}_\kappa t}$: Since $|w_1-w_2|\leqslant C\kappa^2$ by
assumption (\ref{G_23}, \ref{G_24}) and $|w_1|=e^{\Im \lambda_\kappa t}$ we get
\begin{equation*}
|\lambda_\kappa t-\widetilde{\lambda}_\kappa t|
\leqslant
Ce^{-\min(\Im \lambda_\kappa,\Im \widetilde{\lambda}_\kappa)t}\kappa^2\,.
\end{equation*}
Though the exponent is increasing in $t\geqslant 0$ (note that $\Im \lambda_\kappa,\,\Im \widetilde{\lambda}_\kappa \leqslant 0$ by
assumption), the r.h.s. remains uniformly small for $t\leqslant t_0=(\max\{|\Im \lambda_\kappa|,|\Im
\widetilde{\lambda}_\kappa|\})^{-1}$. There the proviso on the modulus can be omitted from the l.h.s. and we obtain for
$t=t_0$
\[|\lambda_\kappa-\widetilde{\lambda}_\kappa|\leqslant C\kappa^{2}
\max\bigl(|\Im \lambda_\kappa|,|\Im \widetilde{\lambda}_\kappa|\bigr)\,.\]
This is the claim (\ref {G_25}) if the maximum is given by $|\Im \lambda_\kappa|$;
in the other case it implies
$|\Im \lambda_\kappa|\geqslant |\Im \widetilde{\lambda}_\kappa|(1-C\kappa^2)$ by the triangle
inequality, and hence again (\ref {G_25}). To prove (\ref{w1w2}) we may
assume $|w_1|>|w_2|$ by symmetry and reduce the claim by homogeneity to
$|\log w|\le[(2+\pi)/2]|w-1|$ for $|w|\geqslant 1$. This follows from
\[|\log w|\leqslant \log|w|+|\arg w|\leqslant |w|-1+\frac{\pi}{2}|w-1|
\leqslant \frac{2+\pi}{2}|w-1|\,,\]
where the estimate on $\log|w|$ is by concavity. That on $\theta=\arg w$
is by the law of cosines
\begin{equation*}
|w-1|^2=|w|^2+1-2|w|\cos\theta=(|w|-1)^2+2|w|(1-\cos\theta)\geqslant
4\sin^2(\theta/2)
\end{equation*}
and by $|\sin(\theta/2)|\geqslant |\theta|/\pi$ for $|\theta|\leqslant\pi$.
\qed\\

\noindent{\bf Acknowledgments} The first named author gratefully
acknowledges the financial support by the Stefano Franscini Fund. The authors 
thank I.M. Sigal for important suggestions. \\

\bibliography{bib}

\end{document}